
\input harvmac.tex

\noblackbox

\def\sq{{\vbox {\hrule height 0.6pt\hbox{\vrule width 0.6pt\hskip 2.5pt
   \vbox{\vskip 6pt}\hskip 2.5pt \vrule width 0.6pt}\hrule height 0.6pt}}\,}

\def\ie{{\it i.e.}}
\def\half{{\scriptstyle{ 1\over2}}}

\def\apm{\alpha^{\prime}}

\def\cD{{\cal D}}
\def\a{\alpha}
\def\b{\beta}
\def\d{\delta}
\def\g{\gamma}

\def\s{\sigma}

\def\e{\epsilon}
\def\p{\partial}

\lref\bhole{G. Horowitz and A. Strominger, {\it Black Strings and p-Branes},
Santa Barbara preprint UCSBTH-91-06 (Mar. 91).}

\lref\dewit{E.~A.~Bergshoeff and M.~de~Roo, {\it The Quartic Effective
Action of the Heterotic Superstring and Supersymmetry}, Nuc.~Phys.~
{\bf B328} (1989) 439.}

\lref\cfmpetal{C. Callan, D. Friedan, E. Martinec and M. Perry,
{\it Strings in Background Fields}, Nuc.~Phys. {\bf B262} (1985) 593.}

\lref\tseyt{E.~Fradkin and A.~Tseytlin, {\it Quantum String Theory
Effective Actions}, Nuc.~Phys. {\bf B261} (1985) 1.}

\lref\asen{A.~Sen, {\it Equations of Motion for Heterotic String Theory
from Conformal Invariance of Sigma Models}, Phys.~Rev.~Lett. {\bf 55}
(1985) 1846.}

\lref\worldsheet{C. Callan, J. Harvey and A. Strominger, {\it Worldsheet
Approach to Heterotic Solitons}, Nuc. Phys. {\bf B} (1991).}

\lref\worldbrane{C. Callan, J. Harvey and A. Strominger, {\it Worldbrane
Actions for Superstrings}, Nuc.~Phys. {\bf B} (1991).}

\lref\gsanom{M. Green and J. Schwarz, Phys.~Lett. {\bf B151} (1985) 21.}

\lref\monosusy{Something on susy approach to finding monopoles}

\lref\hetsol { A. Strominger, {\it Heterotic Solitons},
Nuc.~Phys.~{\bf B343} (1990) 167.}

\lref\DGHR{ A. Dabholkar, G. Gibbons, J. Harvey and F. R. Ruiz,
{\it Superstrings and Solitons}, Nuc.~Phys.~{\bf B340} (1990) 33.}

\lref\DAHA{ A. Dabholkar and J. Harvey, {\it Nonrenormalization of
the Superstring Tension},  Phys.~Rev.~Lett. {\bf 63} (1989) 719.}

\lref\teinep{R. Nepomechie, Phys.~Rev.~{\bf D31} (1985) 1921;
C.~Teitelboim, Phys.~Lett.~{\bf B176} (1986) 69.}

\lref\HLP{ J. Hughes, J. Liu and J. Polchinski,  Phys.~Lett.~{\bf B180}
(1986) 370. }

\lref\howpaptwo{P. S. Howe and G. Papadopoulos,
{\it Finiteness of (4,q) Models with
Torsion}, Nucl.~Phys.~{\bf B289} (1987) 264; {\it Further Remarks
on the Geometry of Two-Dimensional Non-linear $\sigma$ Models},
Class.~Quant.~Grav.~{\bf 5} (1988) 1647. }

\lref\vnew{P. van Nieuwenhuizen and de Wit, {\it Rigidly and
Locally Supersymmetric Two-Dimensional Nonlinear Sigma Models with Torsion},
Nucl. Phys. {\bf B312} (1989) 58.}

\lref\BDHM {T. Banks, L. J. Dixon, D. Friedan and E. Martinec,
{\it Phenomenology and Conformal Field Theory or, Can String Theory
Predict the Weak Mixing Angle?}, Nucl.~Phys.~{\bf B299} (1988) 613.}

\lref\BD{ T. Banks and L. Dixon, {\it Constraints on String Vacua with
Space-time Supersymmetry}, Nucl. Phys. {\bf B307} (1988) 93. }

\lref\eguchi{See e.g. T. Eguchi, H. Ooguri, A. Taormina and S. K. Yang, { \it
Superconformal Algebras and String Compactification on Manifolds
with $SU(N)$ Holonomy}, Nucl. Phys. {\bf B315} (1987) 193 \semi
N. Seiberg, {\it Observations on the Moduli Space of Superconformal
Field Theories}, Nucl. Phys. {\bf B303} (1988) 286. }

\lref\hetsigref{A review by C. Callan and L. Thorlacius can be found in {\it
Particles, Strings and Supernovae}, Vol. 2, eds. A Jevicki and C.-I. Tan,
World Scientific (1989).}

\lref\alfr{L. Alvarez-Gaume and D. Z. Freedman, {\it Geometrical Structure
and Ultraviolet Finiteness in the Supersymmetric Sigma Model},
Comm. Math. Phys. {\bf 80}
443 (1981). }

\lref\BS{T. Banks and N. Seiberg, {\it Nonperturbative Infinities},
Nucl. Phys. {\bf B273} (1986) 157.}

\lref\bachas{I. Antoniades, C. Bachas, J. Ellis and D. Nanopoulos, {\it
Cosmological String Theories and Discrete Inflation},
Phys.~Lett. B{\bf 211} (1988) 393; Nuc.~Phys. B{\bf 328} (1989) 117.}

\lref\hrs{G. Horowitz, A. Strominger and S. J. Rey, unpublished.}

\lref\rusref{V. Dotsenko and V. Fateev, {\it Conformal Algebra and
Multipoint Correlation Functions in Two-Dimensional Statistical Models},
Nucl. Phys. {\bf B240} (1984) 312.}

\lref\ramzi{R. Khuri, {\it Some Instanton Solutions in String Theory},
Princeton preprint PUPT-1219 (1990).}

\lref\ryanetal{R.~Rohm, Phys.~Rev. {\bf D32} (1985) 2849;
R. W. Allen, I. Jack, D. R. T. Jones, {\it Chiral $\sigma$ Models
and the Dilaton $\beta$ Function}, Z. Phys. {\bf C41} (1988) 323.}

\lref\sevrin{A. Sevrin, W. Troost and A. van Proeyen, {\it
Superconformal Algebras
in Two Dimensions with N=4}, Phys. Lett. {\bf B208} (1988) 447.}

\lref\sjrey{S.J. Rey, {\it Axionic String Instantons and Their Low-Energy
Implications}, UCSB preprint  UCSB-TH-89-23, (1989) and {\it
Confining Phase of Superstrings and Axionic Strings}, Phys.~Rev.~{\bf D43}
(1991) 439.}

\lref\RW{ R. Rohm and E. Witten, {\it The Anti-Symmetric Tensor Field in
Superstring Theory}, Ann. Phys. {\bf 170}, 454 (1986). }

\lref\withemil{The discussion which follows was developed in collaboration
with E. Martinec.}

\lref\Hsigref{E.~Braaten and T.~Curtright and C.~Zachos,
{\it Torsion and Geometrostasis in Nonlinear Sigma Models},
Nuc.~Phys.~{\bf B260} (1985) 630.}

\lref\alfr{L. Alvarez-Gaume and D. Z. Freedman, {\it Geometrical Structure
and Ultraviolet Finiteness in the Supersymmetric Sigma Model},
Comm. Math. Phys. {\bf 80}
443 (1981). }

\lref\vnew{P. van Nieuwenhuizen and de Wit, {\it Rigidly and
Locally Supersymmetric
Two-Dimensional Nonlinear Sigma Models with Torsion}, Nucl. Phys. {\bf B312}
(1989) 58.}

\lref\ghr{S. J. Gates, C. M. Hull and M. Rocek, {\it Twisted Multiplets
and New Supersymmetric Nonlinear Sigma Models}, Nucl. Phys. B{\bf 248}
(1984) 157.}

\lref\candelas{P. Candelas, {\it Lectures on Complex Manifolds} in
{\it Superstrings
and Grand Unification, Proceedings of the Winter School on High Energy Physics,
Puri, India, 1988}, T. Pradhan ed., World Scientific Publishers.}

\lref\wzwpap{E. Witten, Comm.~Math.~Phys. {\bf 92} (1984) 455. }

\lref\thooft{See {\it e.g.} R.~Rajaraman {\it Solitons and Instantons},
North Holland, 1982.}

\lref\matmod{Some matrix model review}

\Title{\vbox{\baselineskip12pt\hbox{PUPT-1278}}}
{Instantons and Solitons in Heterotic String Theory}

\baselineskip=12pt
\bigskip
\centerline{Curtis G. Callan, Jr.}
\centerline{\it Department of Physics, Princeton University}
\centerline{\it Princeton, NJ 08544}
\centerline{\it Internet: cgc@pupphy.princeton.edu}

\bigskip
\centerline{\bf Abstract}
This is a transcript of lectures given at the Sixth Jorge Andre Swieca
Summer School in Theoretical Physics. The subject of these lectures is
soliton solutions of string theory. We construct
a class of exact conformal field theories possessing a spacetime soliton
or instanton interpretation and present a preliminary discussion of their
physical properties.

\Date{June 1991}

\newsec{Introduction and Motivation}

For the purposes of these lectures we are going to assume that string theory
is identical to two-dimensional conformal field theory. As is well-known,
conformal field theories of the right central charge and field content
describe solutions of the {\it classical} equations of motion of string
field theory. To obtain true quantum physics one has to sum over higher-genus
worldsheets, an important, but difficult, task which we will not attempt here.

Static solutions of the classical equations of motion of conventional field
theory can represent either candidate vacuum states, solitons or instantons.
The soliton and instanton solutions are distinguished from a vacuum by
being non-invariant to translations and boosts and by having a finite mass
(which goes to infinity as the coupling goes to zero). Vacuum solutions
may have a non-zero energy {\it density}, but this must be interpreted as
a cosmological constant rather than a mass. In short, soliton solutions
identify new physical particles which do not appear in the spectrum of small
fluctuations about the vacuum and may have exotic quantum numbers.
The magnetic monopole solution of spontaneously broken gauge theory is
a classic example. An important development in the study of ordinary quantum
field theory was the demonstration that solitons and their associated
conserved charges survive quantization, and are not just artifacts of the
classical analysis. It is obviously important to repeat this development
for string theory: essentially stringy physics could lead to even more exotic
types of soliton or it could lead to weak instability of the old ones. Either
outcome could have important phenomenological consequences.

Most of what we have learned about string theory so far comes from conformal
field theories with a vacuum interpretation. In order to make contact with
reality, it is necessary to compactify six of the ten spacetime dimensions
of superstring theory and the most carefully studied nontrivial conformal
field theories have precisely this interpretation. The geometry of the six
compactified dimensions may be very complicated, but the remaining four
dimensions are perfectly flat, and the compactification theory manifests
no meaningful energy or mass: it is a vacuum.

To address the soliton issue then, it will be necessary to find conformal field
theories describing localized energy density embedded in asymptotically
flat spacetime (that is, we want to find {\it non}-compactifications!)
Once we find such a solution, we need to develop appropriate conformal field
theory methods for studying its non-vacuum properties, most notably its mass.
As we shall see, these are difficult problems, but some progress has been
made on the first issue.

There are two ways to proceed in the study of this problem. The first is the
"strings in background field" method \refs{\cfmpetal,\tseyt,\asen},
in which the usual spacetime
fields (graviton, dilaton and antisymmetric tensor) appear as coupling
constant functions in a worldsheet nonlinear sigma model and the spacetime
equations of motion for these fields arise from the condition that the
conformal invariance beta functions should vanish. The beta functions are
computed as an expansion in powers of the string tension $\apm$ and, in the
leading approximation, yield standard Einstein-Yang-Mills-like spacetime
equations of motion. Not-so-standard stringy effects arise from higher-order
corrections. It is conceptually straightforward to look for asymptotically
flat solutions of these field equations and most previous attempts to study
string solitons have taken this approach. The problem is that the $\apm$
expansion is only valid if curvatures are everywhere small, a condition
which is not met in many interesting solutions. The second way to proceed
is to use purely algebraic methods to generate exact conformal field theories
in the hope that it will be possible to generate some that have a solitonic
spacetime interpretation. This approach is nonperturbative and perfectly
capable of handling cases with strong curvature, but no useful exact theories
have been found in this way.

In these lectures we will tackle the string soliton problem by a hybrid
of the above two approaches. We will first construct a soliton solution
of the lowest-order spacetime equations and we will then use supersymmetry
arguments to show that this solution gets no corrections to any order in
$\apm$.
Finally, we will use purely algebraic methods to discuss a limiting case
of our general solution and to argue that no nonperturbative effects have
been missed. At the end we will draw conclusions and make some proposals
for further development. The work on which these lectures are based has been
done in collaboration with J. Harvey and A. Strominger and is reported in
\refs{\worldsheet,\worldbrane}. The presentation given here is a pedagogical
elaboration of those references and will, in places, follow them quite closely.

\newsec{Soliton Solutions: Spacetime Approach}

Let us first discuss the problem of finding string solitons via the "strings
in background fields" spacetime approach. The beta functions for strings
propagating in a background of massless fields are the equations of motion
of a certain master spacetime action which can be computed as an expansion
in the string tension $\apm$. For the heterotic string, the leading terms in
this action are identical to the $D=10$, $N=1$ supergravity and super
Yang-Mills action. The bosonic part of this action reads
\eqn\action{S = {1 \over 2 \kappa^2}\int d^{10}x \sqrt{g} e^{-2 \phi} \left(
                R + 4(\nabla \phi)^2 - {1 \over 3} H^2 -
                {\apm \over 30} {\rm Tr} F^2 \right), }
where the three-form antisymmetric tensor field strength is related to the
two-form potential by the familiar anomaly equation \gsanom
\eqn\anom{H=d\wedge B +\apm\left(\omega^L_3(\Omega_-)-{1\over 30}
			\Omega^{YM}_3(A)\right)+\ldots}
(where $\omega_3$ is the Chern-Simons three-form) so that
\eqn\curlh{d\wedge H=\apm (trR \wedge R- {1 \over 30}Tr F \wedge F).}
The trace is conventionally normalized so that
$TrF \wedge F = \sum_i F^i \wedge F^i$ with $i$ an adjoint gauge group index.
An important, and potentially confusing, point is that the connection
$\Omega_\pm$ appearing in \anom\ is a non-Riemannian connection related
to the usual spin connection $\omega$ by
\eqn\connect{ \Omega_{\pm M}^{AB} = \omega_M^{~AB} \pm  H_M^{~AB}. }
Since the antisymmetric tensor field plays a crucial role in all of our
solutions, this subtlety will be crucial.

Rather than directly solve the equations of motion for this action, it is
much more convenient to look for bosonic backgrounds which are annihilated by
{\it some} of the N=1 supersymmetry transformations (only the vacuum is
annihilated by all the the supersymmetries). This is a fairly standard trick
which has been applied to the magnetic monopole problem and to
string problems in \refs{\hetsol,\DGHR}. It is highly nontrivial that any such
solutions can be found at all, but if they can, they are automatically
solutions of the usual equations of motion. The Fermi field
supersymmetry transformation laws which follow from \action\ are
\eqn\susyvar{ \eqalign{\delta \chi & = F_{MN} \gamma^{MN} \epsilon \cr
                       \delta \lambda & = (\gamma^{M} \partial_M \phi -
                       {1 \over 6} H_{MNP}\gamma^{MNP}) \epsilon  \cr
                        \delta \psi_M & = (\partial_M +
                                           {1 \over 4}\Omega_{-M}^{AB}
                                          \gamma_{AB}) \epsilon,   \cr }}
and it is apparent that to find backgrounds for which all of \susyvar\
vanish, it is only necessary to solve first-order equations, rather than
the more complicated second-order equations which follow from varying
the action. We will shortly construct a simple ansatz for the bosonic fields
which does just this.

First, however, we have to specify what type of soliton we are hoping to
construct. In the standard four-dimensional context, we are acquainted
with solitonic solutions of many dimensionalities: Instantons are localized
at a point in Euclidean spacetime and trace out a zero-dimensional worldsheet;
magnetic monopoles are localized at a point in three-dimensional space and
trace out a one-dimensional worldsheet; cosmic strings are localized on a
line in three-dimensional space and trace out a two-dimensional worldsheet and
so on. Generically, we call a soliton whose instantaneous time slice has
p-dimensional extension a {\it p-brane}. In string theory, since the
space-time dimension is ten, we could in principle find solutions with p
anywhere between zero and nine. In these lectures, we will study solutions
with $p=5$: five-branes. They are of particular interest because, as was
shown some time ago by Teitelboim and Nepomechie \teinep\ , five-branes are
dual to fundamental strings in ten dimensions in much the same way that
magnetic
monopoles are dual to electric charge in four dimensions. Dirac's monopole
argument has only to be modified by replacing the Maxwell field $A_\mu$ by the
antisymmetric tensor field $B_{\mu\nu}$ and changing the dimensionality of
spacetime. The duality argument doesn't guarantee that the dual objects
actually exist, of course, but Strominger has shown that, at least
perturbatively, the heterotic string does have soliton solutions
with five-brane structure \hetsol. In these lectures we will concentrate
on constructing five-brane solitons: they are in some sense maximally "stringy"
and, more to the point, many of them are amenable to exact conformal field
theory analysis. How such objects would manifest themselves phenomenologically
is an interesting question which depends on the details of compactification
down to four dimensions. This question has yet to be studied in any detail,
and we will have little to say about it here.

Let us attempt to construct a five-brane solution to \susyvar .
The supersymmetry
variations are determined by a positive chirality Majorana-Weyl $SO(9,1)$
spinor $\e$. Because of the five-brane structure, it is useful to note that
$\e$ decomposes under $SO(9,1)\supset SO(5,1)\otimes SO(4)$ as
\eqn\decomp{ 16 \to (4_+,2_+)\oplus  (4_-,2_-)}
where the $\pm$ subscripts denote the chirality of the representations.
Denote world indices of the four-dimensional space transverse to the fivebrane
by $\mu, \nu = 6 \ldots 9$ and the corresponding tangent space indices by
$m,n =6 \ldots 9$. We assume that no fields depend on the longitudinal
coordinates (those with indices $\mu = 0 \ldots 5$) and that the nontrivial
tensor fields in the solution have only transverse indices. Then the
gamma matrix terms in \susyvar\ are sensitive only to the $SO(4)$ part of $\e$
and, in particular, to its $SO(4)$ chirality.

One immediately sees how to  make the gaugino variation vanish (in what follows
we treat $\e$ as an $SO(4)$ spinor and let all indices be four-dimensional):
As a consequence of the four-dimensional gamma-matrix identity
$\g^{mn}\e_\pm= \mp\half\e^{mnrs}\g^{rs}\e_\pm$ one has
$F_{mn}\g^{mn}\e_\pm= \mp\widetilde F_{mn}\g^{mn}\e_\pm$, where the dual field
strength is defined by $\tilde F_{mn}=\half\e_{mnrs}F^{rs}$. Therefore,
$\delta\chi$ vanishes if
	$F_{mn}=\pm\tilde F_{mn}$ and $\e=(4_\pm,2_\pm)$
which is to say that if the gauge field is taken to be an {\it instanton}, then
$\delta\chi$ vanishes for all supersymmetries with positive $SO(4)$ chirality!

To deal with the other supersymmetry variations, we must adopt an {\it ansatz}
for the non-trivial behavior of the metric and antisymmetric tensor fields
in the four dimensions transverse to the five-brane (the specific form is
inspired by the work of Dabholkar and Harvey \DAHA\ on string-like solitons).
For the metric tensor we write
\eqn\metric{g_{\mu\nu}=e^{-2\phi}\delta_{\mu\nu}\qquad \mu~,\nu=6\ldots 9}
and for the antisymmetric tensor field strength
\eqn\ast{H_{\mu\nu\lambda}=
		\sqrt{g_4}\e_{\mu\nu\lambda\sigma}\partial^\sigma\phi
		=e^{-2\phi}\e_{\mu\nu\lambda\sigma}\partial_\sigma\phi~}
where $\phi$ is to be identified with the dilaton field. With this ansatz
and the rather obvious vierbein choice $e^m_\mu=\delta^m_\mu~e^{-\phi}$,
we can also calculate the generalized spin connections \connect\ which
appear in \susyvar\ and \anom :
\eqn\consatz{{\Omega_\pm}^{mn}_\mu=\delta_{m\mu}\p_n\phi
	-\delta_{n\mu}\p_m\phi\pm\e_{\mu mnp}\p_p\phi~.}

Now consider the $\delta\lambda$ term in \susyvar . Because of the ansatz,
both terms are linear in $\p\phi$. By standard four-dimensional gamma-matrix
algebra, the relative sign of the two terms is proportional to the $SO(4)$
chirality of the spinor $\e$. We have chosen the sign and normalization
of the ansatz for $H$ so that $\delta\lambda$ vanishes for $\e\in (4_+,2_+)$.
Finally, consider the gravitino variation in \susyvar . A crucial fact,
following from \consatz\ , is that while $\Omega_\pm$ would in general be
an $SO(4)$ connection, with the chosen ansatz it is actually pure $SU(2)$.
To be precise,
\eqn\sutwo{{\Omega_\pm}^{mn}_\mu\g^{mn}\e_\eta=
	(\g^{\mu p}\p_p\phi)(1\mp\eta)\e_\eta~,}
so that $\Omega_\pm$ annihilates the $(4_\pm,2_\pm)$ spinor. Since \susyvar\
involves only $\Omega_+$, it suffices to take $\e$ to be a {\it constant}
$(4_+,2_+)$ spinor to make the gravitino variation vanish.

Putting all this together, we see that if we choose the gauge field to be any
instanton and fix the metric and antisymmetric tensor in terms of the dilaton
according to the above ansatz, then the state is annihilated by all
supersymmetry variations generated by a spacetime constant $(4_+,2_+)$ spinor.
Thus, {\it half} of the supersymmetries are unbroken, and the other half,
by standard reasoning \HLP\ , will be associated with fermionic zero-modes
bound to the soliton.

The one unresolved question concerns the functional form of the dilaton field.
Notice that the ansatz for the antisymmetric tensor was given in terms of
its three-form field strength $H_{mnp}$, rather than its two-form potential
$B_{mn}$. This is potentially inconsistent, since the curl of the field
strength must satisfy the anomalous Bianchi identity \curlh\ . Within the
ansatz \ast\ , the curl of $H$ is given by
\eqn\dhsatz{d\wedge H=
    {1\over 4!}\e_{rstu}\p_r\{e^{-2\phi}\e_{stuv}\p_v\phi\}\sim\sq e^{-2\phi}}
(where $\sq$ is just the flat Laplacian) and one can thus, in principle,
solve \curlh\ for $\phi$. The slight problem with this approach is that
\curlh\ is only the leading order in $\apm$ approximation to the true anomaly
and the best we can hope to do is to construct solutions
as a power series in $\apm$. Since our goal is to find exact
solutions, we adopt the different strategy of looking for special
backgrounds where the $R\wedge R-F\wedge F$ anomaly on the {\it r.h.s} of
\curlh\ {\it cancels}. If that is possible,
the equation for $d\wedge H$ becomes $\sq e^{-2\phi}=0$, an equation which
can be solved once and for all with no expansion in powers of $\apm$.
The cancellation of the anomaly of course means that the underlying sigma
model has been made effectively left-right symmetric, a property
which will play a key role in the proof that our solutions are exact in $\apm$.
It remains to show that desired cancellation can, in fact, be achieved.

What is required, according to \curlh\ , is that the curvature $R(\Omega_-)$
should cancel against the instanton Yang-Mills field $F$. We will take
the instanton to be embedded in an $SU(2)$ subgroup of the gauge group
(this is always the lowest-action instanton), so what is needed is that
$\Omega_-$ be a self-dual $SU(2)$ connection. The $SU(2)$ condition we already
know to be met, so the only issue is self-duality. Given the special ansatz
and coordinate system of \metric\ , it is easy to calculate the
curvature of $\Omega_\pm$:
\eqn\omcurv{\eqalign{ R(\Omega_\pm)^{[mn]}_{\mu\nu}=&
	\delta_{n\nu}\nabla_m\nabla_\nu\phi
		-\delta_{m\mu}\nabla_m\nabla_\mu\phi
			-\delta_{m\nu}\nabla_n\nabla_\mu\phi
				+\delta_{m\mu}\nabla_n\nabla_\nu\phi \cr
	&\pm\e_{\mu mn\alpha}\nabla_\alpha\nabla_\mu\phi
		\mp\e_{\nu mn\alpha}\nabla_\alpha\nabla_\nu\phi~~, }}
where
\eqn\phideriv{\eqalign{ \nabla_\mu\nabla_\nu \phi=&
	\p_\mu\p_\nu\phi-\delta_{\mu\nu}\p_m\p_m\phi+2\p_\mu\p_\nu\phi~~,\cr
	\nabla^2\phi =  & 2e^{2\phi}\,\sq e^{-2\phi}~~.  }}
It is then a trivial arithmetical exercise to show that, in four dimensions
and under the condition that $\delta^{mn}\nabla_m\nabla_n\phi =0$,
$\Omega$ is self-dual:
\eqn\omslfdl{R(\Omega_\pm)^{[mn]}_{\mu\nu}=
  \mp\half\e_{\mu\nu\lambda\sigma} R(\Omega_\pm)^{[mn]}_{\lambda\sigma}~.}
Since a self-dual $SU(2)$ connection is an instanton connection, it will
be possible to choose a gauge instanton which exactly matches the
"metric" instanton $\Omega_-$ and makes the {\it r.h.s} of \curlh\
vanish, thus making the whole solution self-consistent.
%
%
In the next section, we will explore the qualitative properties of the
solutions which we have constructed in the above rather roundabout manner.

A feature of the above development which could cause confusion is the intricate
interplay of the two non-Riemannian connections $\Omega_{(\pm)}$. To refresh
the reader's memory, we will summarize the essentials of this phenomenon
(we denote the $(4_+,2_+)$ spinor by $\e_+$): The gravitino supersymmetry
variation equation boils down to $\Omega_{+\mu}^{[ab]}\g^{[ab]}\e_+=0$
which in turn implies that ${R_{\mu\nu}(\Omega_+)}^{ab}\g^{ab}\e_+=0$.
The index-pair interchange symmetry for a non-Riemannian connection, which
reads
$R(\Omega_+)_{ab,cd}=R(\Omega_-)_{cd,ab}$, allows us to convert
the previous condition for $\e_+$ to
${R_{\mu\nu}(\Omega_-)}^{ab}\g^{\mu\nu}\e_+=0$.
If we then make the identification
${F_{\mu\nu}}^{[ab]}\sim {R_{\mu\nu}(\Omega_-)}^{[ab]}$, we see that
we have reproduced the gaugino supersymmetry variation equation. This
is simply to emphasize that, because of the crucial role of the antisymmetric
tensor in these solutions, the precise way in which the $\Omega_+$'s and
$\Omega_-$'s appear in the various equations we deal with is tightly
constrained and quite critical.

\newsec{Development and Interpretation of the Solutions}

Now we will work out the geometry and physical interpretation of the solution
described in the previous section. To recapitulate, we have found the following
solution of the low-energy spacetime effective action of the heterotic string:
\eqn\recap{\eqalign{
	ds^2=&e^{-2\phi(x)}\delta_{\mu\nu}dx^\mu dx^\nu+
		\eta_{\alpha\beta}dy^\alpha dy^\beta \cr
        {H_\mu}^{mn}=&\e_{\mu mnp}\p_p\phi\cr
     H_{\mu\nu\lambda}=&-\half\e_{\mu\nu\lambda\sigma}\p_\sigma e^{-2\phi}\cr
	{F_{\mu\nu}}^{[mn]}=&\hbox{$\widetilde F_{\mu\nu}$}^{[mn]}=
                {R_{\mu\nu}(\Omega_-)}^{[mn]} ~,}}
where $\mu\nu =6\ldots 9$ , $\alpha\beta =0\ldots 5$ and $\eta_{\alpha\beta}$
is the Minkowski metric. The last equation expresses the fact that the
gauge field is a self-dual instanton with moduli chosen so that it coincides
(up to gauge transformations of course) with the curvature of the generalized
connection of the theory. The consistency condition for all this is just
$\sq e^{-2\phi}=0$.

The solution of the consistency condition on $\phi$ is just a constant
plus a sum of poles:
\eqn\phisol{e^{-2\phi}=e^{-2\phi_0}+\sum_{i=1}^N{Q_i\over (x-x_i)^2}}
The constant term is fixed by the (arbitrary) asymptotic value of the dilaton
field, $\phi_0$~. In string theory, $e^{-\phi}$ is identified with the local
value of the string loop coupling constant, $g_{str}$. For the solution
described by \phisol\ , $g_{str}$ goes to a constant at spatial infinity and
goes to infinity at the locations of the poles! We shall worry about the
physical interpretation of this fact in due course. Now, the metric of our
solution is conformally flat with conformal factor given by \phisol\ . Since
$\phi$ goes to a constant at infinity, the geometry is asymptotically flat,
which is precisely what we want for a soliton interpretation.
In the neighborhood of a singularity, we can replace
$e^{-2\phi}$ by a simple pole $Q/r^2$ and obtain the approximate line element
\eqn\wrmmet{\eqalign{
ds^2\sim & {Q\over r^2}(dr^2 + r^2 d\Omega^2_3) \cr
        =& dt^2 + Q d\Omega^2_3 }}
where $d\Omega^2_3$ is the line element on the unit three-sphere and we have
introduced a new radial coordinate $t=\sqrt{Q}log(r/\sqrt{Q})$. This expression
becomes more and more accurate as $t\to -\infty$.
In this same limit, the other fields are given by
\eqn\wrmfld{\phi = t/\sqrt{Q} \quad  H=Q\e_3 ~,}
where $\e_3$ is the volume form on the three sphere. In Sect. 5 we will see
that the linear behavior of the dilaton field plays a crucial role in
the underlying exact conformal field theory. The geometry
described by \wrmmet\ is a cylinder whose cross-section is a three-sphere of
constant area $2\pi^2 Q$. The global geometry is that of a collection of
semi-infinite cylinders, or wormholes (one for each pole in $e^{-2\phi}$),
glued into asymptotically flat four-dimensional space. The wormholes are
semi-infinite since the approximation of \wrmmet\ becomes better and better
as $t\to -\infty$ and breaks down as $t\to +\infty$. It is these wormholes
which we propose to interpret as solitons.

A further crucial fact is that the residues, $Q$, are quantized.
Consider an $S_3$ which surrounds a single pole, of residue Q,
in $e^{-2\phi}$. The net flux of $H$ through this $S_3$ is entirely due
to the enclosed pole and can easily be calculated:
\eqn\Hflux{\eqalign{
H_{ijk}=&-\half\e_{ijkl}\p_le^{-2\phi}=Q\e_{ijkl}\p_lx^l/ x^4 \cr
\int_{S_3}H=&2\pi^2 Q}}
The flux of $H$ through the $S_3$ at in finity is thus proportional to
the sum of all the residues. By a familiar cohomology argument, however,
the flux of $H$ through any $S_3$ must be an integer multiple of some
basic unit \wzwpap . The point is that, if the flux of $H$ is non-zero, then
there cannot be a unique two-form potential $B$ covering the whole sphere. The
best one can do is to have two sections, $B^\pm$, covering the upper and
lower halves of the $S_3$ and related to each other by a gauge transformation
in an overlap region which is topologically an $S_2$. Since the sigma model
action involves $B$, not $H$, the non-uniqueness of $B$ could lead to an
ill-defined sigma model path integral. It is possible to
show that, with our definitions of the sigma model action, this danger is
avoided if and only if the flux of $H$ is an integral multiple of $\apm$. The
details of this argument can be found in \wzwpap~. The consequence for us
is that the residues of the poles in $e^{-2\phi}$ are discretely quantized:
$Q_i=n_i\apm$. As a result, the cross-sectional areas of the individual
wormholes are quantized in units of $2\pi^2\apm$ and there is thus a minimal
transverse scale size of the fivebrane. (This fact may be useful in future
attempts to quantize the transverse fluctuations of the fivebrane.)

Finally, we want to characterize the instanton component of this solution.
The key point is that, when the dilaton field satisfies $\sq e^{-2\phi}=0$,
we can construct a self-dual $SU(2)$ connection out of the scalar field
$\phi$ and we want to identify this connection with the gauge instanton.
But there is a well-known ansatz \thooft\ for constructing multi-instanton
solutions which has precisely this character:
if we define an SU(2) connection by
\eqn\ansatz{
A_\mu(x)=\bar\Sigma_{\mu\nu}\p_\nu log f\qquad
        \bar\Sigma_{\mu\nu}=\half\bar\eta_{\mu\nu i}\sigma^i}
where $\bar\eta_{\mu\nu i}$ is 'tHooft's anti-self-dual symbol,
then the condition of self-duality of the connection reduces to $\sq f=0$,
an equation which has the general solution
\eqn\fsoln{f=1+\sum^N_{i=1}{\rho^2_i\over (x-x_i)^2}~}
(we normalize the behavior at infinity to unity since $A_\mu$ is invariant
to rescaling $f$ by an overall constant). An important point is that the total
instanton number of the solution built on $f$ is $N$, the number of poles.
The gauge potential which follows from taking $f$ to have a
single pole can be shown to be
\eqn\instn{ A_\mu=-2\rho^2\bar\Sigma_{\mu\nu}{x^\nu\over x^2(x^2+\rho^2)}~,}
an expression which one immediately recognizes as the singular gauge instanton
of scale size $\rho$ centered at $x=0$. The only way this can match our
construction of a self-dual generalized connection is if we make the
identification
\eqn\ident{f(x)=e^{2\phi_0}e^{-2\phi}=
                        1+\sum^N_{i=1}{e^{2\phi_0}Q_i\over (x-x_i)^2}~.}
Thus, given the solution \phisol\ for the dilaton field, we can assert that
the associated instanton has instanton number $N$, with instantons of scale
size $e^{2\phi_0}Q_i$ localized at positions $x_i$. Since the $Q_i$ are
quantized, so are the instanton scale sizes. The only free parameters (moduli)
are the $4N$ center locations of the instantons. In ten dimensions, the
multiple instanton solution corresponds to multiple fivebranes with the
locations in the transverse four-dimensional space of the individual fivebranes
given by the center coordinates of the individual instantons.

An important fact about the solution we have just constructed is that it is
not perturbative in $\apm$. As we saw in the discussion following \recap ,
the wormhole associated with a pole of residue $Q=n\apm$ has a cross-section
which is a sphere of area $2\pi^2Q$ and therefore has curvature $R\sim 1/Q$.
In the perturbative sigma model approach to strings in background fields,
one finds that the sigma model expansion parameter is $\apm R$. In the case
at hand, this becomes $\apm/Q= 1/n$, which is obviously not small for the
elementary fivebrane, which has $n=1$. Since our solution has been constructed
by solving the leading-order-in-$\apm$ beta function equations, ignoring all
higher-order corrections, one can legitimately worry whether it makes any
sense. In the next two sections we will present evidence that all higher-order
corrections to this particular solution actually vanish, and the leading-order
solution is exact. We will briefly discuss other solutions of interest for
which higher-order corrections don't vanish.

There is another perturbation theory issue to bring up here. String theory has
two expansion parameters: the string tension $\apm$ and the string loop
coupling constant $g_{str}\sim e^{-2\phi_0}$. The latter is the quantum
expansion parameter of string theory and, in this paper, we are working to
zeroth-order in an expansion in $g_{str}$. In effect, we are producing an
exact solution, to all orders in $\apm$, of {\it classical} string field
theory. However, as we have already pointed out, our solution has the unusual
feature that $g_{str}$ grows without limit down the throat of a wormhole
so that there is, strictly speaking, no reliable classical limit! Since
virtually nothing is known about non-perturbative-in-$g_{str}$ physics, we
don't know what this means for the ultimate validity of this sort of solution.
Similar issues arise in the matrix model/Liouville theory approach to
two-dimensional quantum gravity, and we hope eventually to gain some insight
from that source (for a review, see the contribution of David Gross to
this School).

Before proceeding to show that our solution is exact (in the sense just
described), we want to briefly describe some inexact, but instructive,
solutions of the basic heterotic field equations. A particularly interesting
possibility (and this was Strominger's original approach to this problem
\hetsol ) is to proceed along the lines of Sect. 2, but to determine
the dilaton by solving the curl equation for $H$ perturbatively in $apm$.
To do this, combine \dhsatz\ and \curlh to give
\eqn\anomeq{\sq e^{-2\phi}\sim\apm (tr R\wedge R-{1\over 30}Tr F\wedge F)~.}
In a perturbative solution, this equation implies that $\p\phi\sim O(\apm)$
which, via \omcurv\ , implies that $R\sim O(\apm)$. Therefore, to
leading order in $\apm$, one is entitled to drop the $R\wedge R$ term in
\anomeq\ . Substituting the explicit gauge field strength for an instanton
of scale size $\rho$, one obtains the following dilaton solution:
\eqn\stromdil{e^{-2\phi}=e^{-2\phi_0}+
                8\apm{(x^2+2\rho^2)\over(x^2+\rho^2)^2}+O({\apm}^2)~.}
The metric and antisymetric tensor fields are built out of this dilaton field
according to the spacetime-supersymmetric ansatz of \metric\ and \ast .
This solution is very different from the previous one, obtained by setting
$d\wedge H=0$: The dilaton field and the metric are everywhere finite and
the topology of the solution is $R_4$ rather than $R_4$ with semi-wormholes
glued in. One can examine higher-order in $\apm$ corrections to the beta
functions and verify that the solution must receive corrections. At the
same time, one can examine higher-order corrections to the supersymmetry
transformations \dewit\ and verify that it is possible to maintain
spacetime supersymmetry in the $\apm$-corrected solution. These solutions
are very interesting in their own right and certainly have a soliton
interpretation. On the other hand, since we do not know how to deal with
the higher-order in $\apm$ corrections in any general manner, we will not
pursue this line of development here.

Another interesting point concerns what happens when we lift the requirement
of spacetime supersymmetry and look for solutions of the beta function
equations rather than the condition that some supersymmetry charges annihilate
the solution. Our solutions have the property that the mass (the ADM mass, to
be precise) per unit fivebrane area is proportional to the axion charge:
$M_5=2\pi{\apm}^{-3}Q$. This equality can be understood via a Bogomolny
bound: any solution of the leading-order field equations with the fivebrane
topology must satisfy the inequality $M_5\ge 2\pi{\apm}^{-3}Q$ and our
solution saturates the inequality. One can easily imagine a process in which
mass, but not axion charge, is increased by sending a dilaton wave down one of
the wormhole throats. Since the wormhole throat is semi-infinite, this wave
need
not be reflected back: It can continue to propagate down the throat forever,
leaving an exterior solution for which $M_5>2\pi{\apm}^{-3}Q$. Such solutions
of the leading-order beta function equations have indeed been found \bhole\
and they resemble the familiar Reissner-Nordstrom sequence of charged black
holes: They have an event horizon and a singularity, but the singularity
retreats to infinity as the mass is decreased to the extremal value that
saturates the Bogomolny bound. Perhaps not surprisingly, the non-extremal
solutions are not annihilated by any spacetime supersymmetries, and we do
not expect to be able to find the corresponding exact conformal field theories.
Nonetheless, the fact that the extremal black hole is under exact string
theory control should eventually allow us to make progress on understanding the
string physics of black holes, Hawking radiation and the like.

In the rest of these lectures, we will pursue the much more limited goal of
showing that our special solution is an exact solution of string theory.

\newsec{Worldsheet Sigma Model Approach}

To show conclusively that a given spacetime configuration is a solution of
string theory, we must show that it derives from an appropriate
superconformal worldsheet sigma model. In this section we will show that
the worldsheet sigma models corresponding to the fivebranes constructed
in section 2 possess extended worldsheet supersymmetry of type (4,4)
The notation derives from
the fact that in a conformal field theory, the left-moving fields (functions
of $z$) and the right-moving fields (functions of $\bar z$) are dynamically
independent. It is therefore possible to have different numbers of right- and
left-moving supercharges $Q^I_\pm$. The general case, referred to as
$(p,q)$ supersymmetry, is described by the algebra
\eqn\extalg{\eqalign{
\{ Q^I_+, Q^J_+\}=&\delta^{IJ}\p_z\qquad I,J=1\ldots p\cr
\{ Q^I_-, Q^J_-\}=&\delta^{IJ}\p_{\bar z}\qquad I,J=1\ldots q\cr
\{ Q^I_+, Q^J_-\}=&0~~.}}
The minimal possibility, corresponding to a generic solution of the heterotic
string, has $(1,0)$ supersymmetry. Any left-right-symmetric, and therefore
non-anomalous theory, will have $(p,p)$ supersymmetry (this is sometimes
referred to as N=p supersymmetry). The maximal possibility
is $(4,4)$ which, it turns out, is what is realized in our fivebrane solution.
We will argue that, in the (4,4) case,
there is a nonrenormalization theorem which makes the lowest-order in $\apm$
solution for the spacetime fields exact. The latter issue is closely related
to the question of finiteness of sigma models with torsion and with extended
supersymmetry \refs{\howpaptwo, \vnew} and the results we find are slightly
at variance with the conventional wisdom, at least as we understand it. We will
comment upon this at the appropriate point.

First we digress to explain why we expect four-fold extended
supersymmetry in this problem. The models of interest to us are structurally
equivalent to a compactification of ten-dimensional spacetime down to six
dimensions: there are six flat dimensions (along the fivebrane) described
by a free field theory and four `compactified' dimensions (transverse to the
fivebrane) described by a nontrivial field theory. The fact that the
`compactified' space is not really compact has no bearing on the supersymmetry
issue. The defining property of all the fivebranes of section 2 is that they
are
annihilated by the generators of a six-dimensional N=1 {\it spacetime}
supersymmetry. That is, they provide a compactification to six
dimensions which maintains N=1 spacetime supersymmetry. Now, it is well-known
that in compactifications to four dimensions, the sigma model describing the
six compactified dimensions must possess (2,0) worldsheet supersymmetry in
order for the theory to possess N=1 four-dimensional spacetime supersymmetry
\BDHM. Roughly speaking, the conserved U(1) current of the (2,0)
superconformal algebra defines a free boson which is used to construct the
spacetime supersymmetry charges. It is also known that, if one wants to impose
N=2 four-dimensional spacetime supersymmetry, the compactification sigma model
must have (4,0) supersymmetry \BD. The conserved SU(2) currents of the (4,0)
superconformal algebra are precisely what are needed to construct the larger
set
of N=2 spacetime supersymmetry charges. Since, by dimensional reduction, N=1
supersymmetry in six dimensions is equivalent to N=2 in four dimensions, the
above line of argument implies that spacetime supersymmetric compactifications
to six dimensions (including our fivebrane) require a compactification sigma
model with at least (4,0) worldsheet supersymmetry. Since our solution is
constructed to cancel the anomaly, it will be left-right symmetric and
therefore automatically of type $(4,4)$.

Now we turn to a study of string sigma models. The generic
sigma model underlying the heterotic string describes the dynamics of D
worldsheet bosons $X^M$ and D right-moving worldsheet fermions $\psi^M_R$
(where D, typically ten, is the dimension of spacetime) plus left-moving
worldsheet fermions $\lambda^a_L$  which lie in a representation of the gauge
group $G$ (typically $SO(32)$ or $E_8\otimes E_8$). The generic Lagrangian for
this sigma model is written in terms of coupling functions $G_{MN}$,
$B_{MN}$ and $A_M$ which eventually get interpreted as spacetime metric,
antisymmetric tensor and Yang-Mills gauge fields. This Lagrangian  has
the explicit form \hetsigref\
\eqn\hetsiglag{\eqalign{{1\over
4\pi\apm}\int d^2\sigma\{ &
       G_{MN}(X)\partial_+X^M\partial_-X^N +
            2   B_{MN}(X)\partial_+X^M\partial_-X^N \cr
+&iG_{MN}\psi^M_R{\cal D}_-\psi^N_R +i\delta_{ab}\lambda^a_L{\cal D}_+
\lambda^b_L
     +\half (F_{MN})_{ab}\psi^M_R\psi^N_R\lambda^a_L\lambda^b_L\}~~}}
where $H=dB$.  In this expression, the covariant derivatives on the
left-moving fermions are defined in terms of the Yang-Mills connection,
while the covariant derivatives on the right-moving fermions are defined in
terms of a non-riemannian connection involving the torsion (which already
appeared in section 2):
\eqn\covderiv{\eqalign{
{\cal D}_- \psi^A_R=&\partial_-\psi^A_R+
    {{\Omega_{-N}}^A}_B \partial_-X^N\psi^B_R, \cr
{\cal D}_+\lambda^a_L=&\partial_+\lambda^a_L+
    {{A_{N}}^a}_b\partial_+X^N\lambda^b_L . }}
As in section 2 we use indices of type $M$ for coordinate space indices,
type $A$ for the tangent space and type $a$ for the gauge group.
An absolutely crucial feature of this action is that the connection appearing
in
the covariant derivative of the right-moving fermions is the generalized
connection $\Omega_-$, {\it not} the Christoffel connection. This action has
a naive $(1,0)$ worldsheet supersymmetry and can be written in terms of (1,0)
superfields. Superconformal invariance is broken by anomalies of various kinds
unless the coupling functions satisfy certain `beta function' conditions
\cfmpetal\ which are equivalent to the spacetime field equations discussed in
section 2.  The dilaton enters these equations in a rather roundabout, but
by now well-understood, way \hetsigref.

To proceed further, we must construct the specific sigma models corresponding
to the fivebrane solutions. For the generic fivebrane, \hetsiglag\
undergoes a split into a nontrivial four-dimensional theory and a
free six-dimensional theory: the sigma model metric (as opposed to the
canonical general relativity metric) then describes a flat six-dimensional
spacetime times four curved dimensions. The right-moving fermions couple
via the kinetic term to the generalized connection $\Omega_-$, which acts only
on the four right-movers lying in the tangent space orthogonal to the
fivebrane. The other six right-movers are free (we momentarily ignore the
four-fermi coupling) so there is a six-four split of the right-movers as well.
The left-moving fermions couple to an instanton gauge field which
may or may not be identified with the {\it other} generalized
connection, $\Omega_+$. In all the cases of interest to us, the gauge
connection
is an instanton connection and acts only in some $SU(2)$ subgroup of the full
gauge group, so that four of the left-movers couple nontrivially, while the
other 28 are free. Finally, the four-fermion interaction term couples together
precisely those left- and right-movers which couple to the nontrivial gauge and
$\Omega_-$ connections and is therefore consistent with the six-four split
defined by the kinetic terms. The remaining variables can be regarded as
defining a heterotic, but free, theory (6 $X$, 6 $\psi_R$ and 28 $\lambda_L$)
living in the six `uncompactified' dimensions along the fivebrane. From now
on, we focus our attention on the nontrivial piece of \hetsiglag\ referring to
the four-dimensional part of the split. For string theory consistency, it must
have a central charge of 6, which would be trivially true if the connections
were all flat, but is far from obvious for a fivebrane.

Now let us further specialize to the sigma model underlying the left-right
symmetric (and therefore non-anomalous) fivebrane solution of section 2. It is
constructed by identifying the gauge connection with the `other' generalized
connection $\Omega_+$ and making that connection self-dual by imposing the
condition $\sq e^{-2\phi}=0$ on the metric conformal factor. The result of
this is that the four bosonic coordinates transverse to the fivebrane and
the four nontrivially-coupled left- and right-moving fermions are governed
by the worldsheet action
\eqn\torsig{\eqalign{{1\over 4\pi\apm}\int
d^2\sigma\{ &
       G_{\mu\nu}(X)\partial_+X^\mu\partial_-X^\nu +
             2  B_{\mu\nu}(X)\partial_+X^\mu \partial_-X^\nu \cr
+&iG_{\mu \nu}\psi^\mu_R{\cal D}_-\psi^\nu_R +iG_{\mu \nu}
\lambda^\mu_L{\cal D}_+\lambda^\nu_L
       +\half R(\Omega_+)_{\mu
\nu \lambda \rho}\psi^\mu_R\psi^\nu_R\lambda^\lambda_L\lambda^\rho_L\} }}
where ${\cal D}_\pm$ are the covariant derivatives built out
of the generalized connections $\Omega_\pm$ . In fact, as long as the $H$
appearing in $\Omega_\pm$ is given by $d\wedge B$, \torsig\ is identical to
the basic left-right symmetric, $(1,1)$ supersymmetric nonlinear sigma
model with torsion \Hsigref\ .
Despite the apparent asymmetry of the coupling of $\lambda_L$ to $\Omega_+$
and $\psi_R$ to $\Omega_-$, the theory nonetheless has an overall left-right
symmetry (under which $B \rightarrow -B$) and is non-anomalous. To exchange the
roles of $\psi_R$ and $\lambda_L$ one has to replace the curvature of
$\Omega_-$ by that of $\Omega_+$. This exchange symmetry property relies on
the non-riemannian relation
\eqn\nonriem{ R(\Omega_+)_{\mu \nu \lambda \rho}=
 R(\Omega_-)_{\lambda \rho \mu \nu}}
which indeed holds for the generalized connection \connect\  when $d\wedge
H=0$.
To summarize, we have shown that the heterotic sigma model describing the
nontrivial four-dimensional geometry of the fivebrane is actually an example
of a left-right symmetric sigma model with at least $(1,1)$ supersymmetry.
As we will now show, it actually has $(4,4)$ worldsheet supersymmetry.

We now turn to the question of extended supersymmetry. The basic worldsheet
supersymmetry of a $(1,1)$ model like \torsig\ is
\eqn\susyone{ \eqalign{
\d X^M=\e_L\psi^M_R+&\e_R\psi^M_L \cr
\d \psi^A_L+[\Omega_{+M}]^{AB}\d X^M\psi^B_L=&\p X^A\e_R+\ldots \cr
\d \psi^A_R+[\Omega_{-M}]^{AB}\d X^M\psi^B_R=&\p X^A\e_L+\ldots }~.}
The worldsheet supersymmetry of the $(1,0)$
model is obtained by dropping the contributions of $\e_R$ and $\psi_L$.
The general structure of a possible second supersymmetry transformation is
\eqn\susytwo{ \eqalign{
\hat\d X^M=\e_Lf_R(X)^M_{~N}\psi^N_R+&\e_Rf_L(X)^M_{~N}\psi^N_L \cr
\hat\d \psi^A_L+[\Omega_{+M}]^{AB}\d X^M\psi^B_L=&
		-f_L(X)^A_{~B}\p X^A\e_R+\ldots \cr
\hat\d \psi^A_R+[\Omega_{-M}]^{AB}\d X^M\psi^B_R=&
		-f_R(X)^A_{~B}\p X^A\e_L+\ldots ~.}}
The function $f$ is normalized and fully defined by the requirements that
$\{\hat\d,\d\}=0$ and that $\hat\d$ anticommute with itself to give ordinary
translations as in \extalg\ . The question is, what conditions must $f$
satisfy in order for $\hat\d$ to be a symmetry and how many of them can
there be?

This question was first addressed in \alfr\ for the case of left-right
symmetric theories without torsion (\ie without an antisymmetric tensor
coupling term). The more complex case of left-right symmetry with torsion
was subsequently dealt with in \refs{\ghr,\howpaptwo,\vnew}. The basic result
is that the pair of tensors $f_{R,L}$ must be complex structures, covariantly
constant with respect to the appropriate connection:
\eqn\cmpstr{\eqalign{
f^2_\pm = & -1 \cr
	{\cal D}^\pm_A {f_\pm}^B_{~C}=& \p_A{f_\pm}^B_{~C}+
		{\Omega^{(\pm)}_{AD}}^B {f_\pm}^D_{~C}-
			{\Omega^{(\pm)}_{AC}}^D {f_\pm}^B_{~D}=0~,}}
where the $\pm$ notation is equivalent to the $L,R$ notation. The tensors in
\cmpstr\ are written in tangent space indices which is why the generalized
spin connections $\Omega^{(\pm)}$ appear in the covariant derivative. The
equation could, of course, also have been written in coordinate indices.
In general, it is not obvious that such a pair of complex structures can be
found, but, if one can, we know that the sigma model actually possesses
$(2,2)$ worldsheet supersymmetry. A further question is whether multiple
pairs $f^{(r)}_\pm$ of such complex structures can be found. If we can find
$p-1$ of them, then the sigma model has $(p,p)$ supersymmetry. It turns out
that the only consistent possibility for multiple complex structures is that
there be three of them \alfr\ and that they satisfy the Clifford algebra
\eqn\clifford{f^{(r)}_\pm f^{(s)}_\pm =-\d_{rs}+\e_{rst}f^{(t)}_\pm ~.}
This corresponds to the case of $(4,4)$ supersymmetry. It is worth noting
that each complex structure leads to a conserved (chiral) current:
\eqn\concur{J_\pm^{(r)}=\psi^A_\pm (f^{(r)}_\pm)_{AB}\psi^B_\pm~.}
This yields a $U(1)$ symmetry in the $(2,2)$ case and an $SU(2)$ symmetry in
the $(4,4)$ case.

The question of left-right asymmetric theories, such as those which underlie
the `non-exact' fivebranes discussed in section 3, is more delicate.
According to \howpaptwo\ , a heterotic sigma model will have $(p,0)$
supersymmetry if there are $p-1$ complex structures $f^{(r)}_+$ which
are covariantly constant under the connection which couples to the right-moving
fermions (those which do not couple to the gauge field) and if the gauge field
(which affects the left-moving fermions) satisfies a condition which reduces,
for a four-dimensional base space, to self-duality. The latter condition
is met for all of the fivebranes of interest to us since they are all built
on instanton gauge fields. Thus, in all cases, the essential issue is the
existence of complex structures.

To count complex structures, we will use the theorem that a complex
structure can be constructed from any covariantly constant spinor \candelas .
We start with a spinor $\eta$ (in our case four-dimensional) of definite
chirality ($\g_5\eta=\pm\eta$, say) and unit normalized ($\eta^\dagger\eta=1$).
Then we define a tensor
\eqn\construct{J_{AB}=-i\eta^\dagger\g_{AB}\eta}
which we will
identify as a complex structure tensor (in tangent space indices and with
indices raised and lowered by the identity metric). It is then
automatic that if the spinor is covariantly constant with respect
to some connection, so is $J_{AB}$. A simple Fierz identity argument,
quite similar to that found on p.52 of \candelas\ , then shows that $J$ squares
to $-1$ ($J_{AB}J_{BC}=-\d_{AC}$) and is indeed a complex structure.

We are now ready to construct the explicit complex structures.
As was explained in the discussion after \sutwo\ , on the fivebrane,
{\it constant} spinors of definite four-dimensional chirality are covariantly
constant. Using the Weyl representation for the four-dimensional gamma
matrices, one has the following solutions of the two covariant constancy
conditions:
\eqn\cstspin{\eqalign{
\cD_\mu(\Omega_+)\e_+=0~&\Rightarrow~\e_+=\pmatrix{\chi\cr 0} \cr
\cD_\mu(\Omega_-)\e_-=0~&\Rightarrow~\e_-=\pmatrix{0\cr \chi} ~,}}
where $\chi$ is {\it any} constant two-spinor (which we might as well unit
normalize). Since there are three parameters needed to specify the general
normalized two-spinor, there should be three independent choices for the
two-spinor $\chi$ and therefore three choices for both $\e_+$ and $\e_-$.
We will define the independent $\chi_r$ ($r=1,2,3$) as those which give
expectation values of the spin operator along the three coordinate axes:
\eqn\expect{ \chi^\dagger_r\sigma^i\chi_r=\d_{ir} .}
This finally leads, with the help of \construct\ , to the following set of
three right- and left-handed complex structures:
\eqn\result{\eqalign{
J^+_1=\pmatrix{i\s_2&0\cr 0&i\s_2} &\quad
										J^-_1=\pmatrix{-i\s_2&0\cr 0&-i\s_2}\cr
J^+_2=\pmatrix{0&1\cr -1&0} &\quad
										J^-_2=\pmatrix{0&-\s_3\cr \s_3&0}\cr
J^+_3=\pmatrix{0&i\s_2\cr i\s_2&0} &\quad
										J^-_3=\pmatrix{0&-\s_1\cr \s_1&0}~.}}
It is trivial to show that the $J^+$ commute with all the $J^-$ and that they
satisfy the Clifford algebra \clifford . These are precisely the conditions
needed to generate $(4,4)$ supersymmetry in a left-right symmetric theory
(or $(4,0)$ supersymmetry in a heterotic theory). The complex structures are
thus extremely simple indeed.

Finally, we come to the questions of finiteness and need for higher-order
in $\apm$ corrections to our solutions. It is rather firmly established
that two-dimensional nonlinear sigma models with $(4,4)$ supersymmetry are
in fact finite. The general proof was given quite some time ago by
Alvarez-Gaume
and Freedman \alfr\ and it consists in showing that no $(4,4)$-invariant
counterterms of the needed dimension can be constructed. If the theory is
finite, the beta-functions get no higher-order corrections and the choice of
background fields which made the beta functions vanish at leading order must
continue to make them vanish at all orders in $\apm$. Confirmation of this
comes from a construction by Gates {\it et.al.} \ghr , using $(2,2)$
superfields, of the most general $(4,4)$-invariant action. The functional
form of the action must satisfy certain conditions in order to have $(4,4)$
supersymmetry and, with hindsight, one can see that the most general solution
of these conditions corresponds precisely to our special multi-fivebrane
solution.

As an aside, we mention that it has been argued that one really only needs
$(4,0)$ supersymmetry to achieve finiteness \howpaptwo . This would apply
to variations on the solution described in Sect.~2 in which, for example,
the gauge instanton scale size did not match the wormhole throat transverse
scale size. In the discussion given earlier in this section,
we recall that the existence and properties of the right-moving complex
structures $f^{(+)}_i$ have nothing to do with the properties of the gauge
field (which governs the left-moving complex structures). So, if we keep the
same metric then we should have the same $f^{(+)}_i$ and thus at least a
$(4,0)$ supersymmetry. While the solution may well exist, the anomalies
probably mean that there will be corrections to the beta functions so that
the theory is not finite, but constructible order by order. This subject has
yet to be explored in any detail.

\newsec{Algebraic CFT Approach}

It is one thing to show that a sigma model is a superconformal field theory, as
we have done in the previous section, and quite another to be able to classify
its primary field content and calculate n-point functions of its vertex
operators. Indeed, in order to answer all the interesting questions about
string
solitons, it would be desirable to have as detailed an algebraic
understanding of the underlying conformal field theory as we already have for,
say, the minimal models. We are far from having such an understanding,
but in this section we will see that useful insight can be gained by studying a
special limit which emphasizes the semi-wormhole throat.

Recall from section 2 that the (four-dimensional part of the) metric
of the symmetric solution has the form
\eqn\mmetric{ds^2=e^{-2\phi}dx^2}
where $dx^2$ is the flat metric on $R^4$ and
\eqn\confact{e^{-2\phi(x)}=e^{-2\phi_0}+\sum_1^n{Q_i\over (x-x_i)^2}~.}
The singularities in $e^{-2\phi}$ are associated with the semi-wormholes.
Taking $n=1$ and the limit $e^{-2\phi_0} \rightarrow 0$ gives
\eqn\wormlim{e^{-2\phi}= {Q\over x^2},}
which is the solution corresponding to the wormhole throat
itself. Using spherical coordinates
centered on the singularity, and defining a logarithmic radial coordinate by
$t=\sqrt{Q}ln \sqrt{x^2/Q}$, the metric, dilaton  and axion
field strength of the throat may be written in the form
\eqn\wormsol{\eqalign{ds^2&= dt^2 + Q d\Omega_3^2, \cr \phi &
=-t  / \sqrt{Q}, \cr H &=-Q\epsilon,} }
where $d\Omega_3^2$ is the line element and $\epsilon$ the volume form
of the unit 3-sphere obeying $\int \epsilon =2\pi^2$. The geometry of the
wormhole is thus a 3-sphere of radius $\sqrt{Q}$ times the open line $R^1$
and the dilaton is linear in the coordinate of the $R^1$.
Remarkably, these metric and antisymmetric tensor fields are such that the
curvatures constructed from the generalized connections, defined in \connect\ ,
are identically zero, reflecting the parallelizability of $S^3$.
The axion charge $Q$ is integrally quantized. So, since $Q$ appears in the
metric, the radius of the $S^3$ is quantized as well.

The sigma model defined by these background fields is an interesting variant of
the Wess-Zumino-Witten model and the underlying conformal field theory can, it
turns out, be analyzed in complete detail. The basic observation along these
lines was made in \bachas\ in the lorentzian context and euclideanized in
\refs{\hrs, \sjrey}: the $S^3$ and the antisymmetric tensor field are
equivalent to the $O(3)$ Wess-Zumino-Witten model of level
\eqn\level{k={Q\over \apm} ,} while the $R^1$ and the linear
dilaton define a Feigin-Fuks-like free field theory with a background charge
induced by the linear dilaton. Both systems are conformal field
theories of known central charges:
\eqn\wormccs{c_{wzw}={3k\over k+2}\qquad c_{ff}=1+{6\over k}~~.}
The shift of the $R^1$ central charge away from unity is a familiar background
charge effect which has been exploited in constructions of the minimal
models \rusref\ and in cosmological solutions \bachas.

For the combined
theory to make sense, the net central charge must be four. Let us for the
the moment consider the bosonic string. If we expand $c_{wzw}$ in powers of
$k^{-1}$ (this corresponds to the usual perturbative expansion in powers of
$\apm$), we see instead that
\eqn\ctot{c_{tot}=c_{wzw}+c_{ff}=4 + O(k^{-2}\sim{\apm}^2) ~~.}
But, we should not have expected to do any better: the
field equations we solved in section 2 to get this solution are
only the leading order in $\apm$ approximation to the full bosonic string
theory field equations and we must expect higher-order corrections to the
fields
and central charges. In fact, this issue can be studied in detail and it
can be shown \ramzi\ that the metric and antisymmetric tensor fields are not
modified and that the only modification of the dilaton is to adjust the
background charge of the $R^1$ ({\it i.e.} the coefficient of the
linear term in $\phi$) so as to maintain $c_{tot}$ exactly equal to four.

While this is quite interesting, we are really interested in the
superstring case. The leading-order-in-$\apm$ metric, dilaton etc.
fields are the same as in the bosonic case (and, because of the
non-renormalization theorems, we
expect no corrections to them) but various fermionic terms are added to the
previous purely bosonic sigma model. The structure is that of the (1,1)
worldsheet supersymmetric sigma model \torsig\ discussed in section 3. There
is still an $S^3\times R^1$ split, but the component theories are
supersymmetrized versions of Wess-Zumino-Witten and Feigin-Fuks. The
Feigin-Fuks theory is still essentially free. In the supersymmetric WZW theory,
the four-fermi terms vanish identically because, as pointed out above,
the generalized curvature vanishes for this background. As a consequence, the
generalized connections are locally pure gauge
and can be eliminated from the fermion
kinetic terms by a gauge rotation of the frame field.
Since the fermions are effectively free, they
make a trivial addition to the  central charges of both the $S^3$ and the
$R^1$ models:
\eqn\fwormcc{c_{wzw}={3 k\over k+2}+{3\over 2}
                             \qquad c_{ff}=1+{6\over k}+{1\over 2}~~.}
There is, however, a small subtlety: the gauge rotation which decouples the
fermions is {\it chiral}, and therefore anomalous, because the left- and
right-moving fermions couple to two {\it different} pure gauge generalized
connections, $\Omega_+$ and $\Omega_-$. The entire effect of this anomaly on
the
central charge turns out to be the replacement in $c_{wzw}$ of $k$ by $k-2$
(the details can be found in \ryanetal) with the result that
\eqn\anomcc{c_{wzw}={3 (k-2)\over k}+{3\over 2}
					\qquad c_{tot}=c_{wzw}+c_{ff}=6~~.}
Six is, of course, exactly the value we want for the central charge. The
remarkable fact is that, in the supersymmetric theory,
the expansion of $c_{wzw}$
in powers of $k^{-1}$ {\it terminates} at first non-trivial order and no
modification of the dilaton field is needed to maintain the desired central
charge of six. These results are consistent with the non-renormalization
theorems discussed in section 4, but are not tied to perturbation theory,
since they derive from exactly-solved conformal field theories. On the other
hand, since the present discussion makes no reference to the (4,4)
supersymmetry which was crucial in proving the perturbative
non-renormalization theorems of section 4, an important element
is still missing.

This is a good point to remind the reader of the hierarchy of superconformal
algebras. Much of what we know about conformal field theory comes from studying
the representation theory of these algebras. The basic N=1 superconformal
algebra is contains an energy-momentum tensor $T(z)$ and its superpartner
$G(z)$. The essential information is contained in the singular terms in their
operator product expansion:
\eqn\algone{\eqalign{
T(z)T(w)=&{c/2\over(z-w)^4}+{2T(w)\over(z-w)^2}+{\p_wT(w)\over(z-w)}+\ldots\cr
T(z)G(w)=&{{3\over 2}G(w)\over(z-w)^2}+{\p_wG(w)\over(z-w)}+\ldots\cr
G(z)G(w)=&{c/6\over(z-w)^3}+{\half T(w)\over(z-w)}+\ldots~.}}
The central charge $c$ is unconstrained. All superstring theories have at
least this much worldsheet supersymmetry. The N=2 superconformal algebras
differ from this by having a conserved current $J(z)$ and {\it two}
supercharges $G^\pm(z)$ distinguished by the value ($\pm 1$) of their charge
with respect to the current $J(z)$. This charge also plays a key role in the
GSO projection which rids the theory of tachyons. The important new algebraic
relations are contained in the operator products
\eqn\algtwo{\eqalign{
J(z)G^\pm(w)=&\pm{G^\pm(w)\over(z-w)}+\ldots\cr
G^\pm(z)G^\pm(w)\sim & 1\cr
G^+(z)G^-(w)=&{c/6\over(z-w)^3}+{\half J(w)\over(z-w)^2}+\ldots~.}}
There is an N=1 subalgebra generated by $T(z)$ and
$G(z)={1\over\sqrt{2}}(G^+(z)+G^-(z))$. The `practical' utility of the N=2
algebra is that the conserved current defines a free field $H$ by the relation
$J(z)=i\sqrt{c\over 3}\p_z H(z)$ and this free field can be used to construct
the N=1 {\it spacetime} supersymmetry charge in a compactification to four
dimensions. Once again, the central charge, $c$, is unconstrained.
One further extension, to four supercharges, turns out to be possible
(and it can be shown \alfr\ that this is the maximal extension). There are
now three conserved currents $J^i$ which generate an $SU(2)$ Kac-Moody algebra
and the supercharges $G^\a(z),\bar G^\a(z)$ are in $I=1/2$ representations
of the conserved $SU(2)$. The relevant operator product expansions are
\eqn\algfour{\eqalign{
T(z)T(w)=&{3k\over(z-w)^4}+{2T(w)\over(z-w)^2}+{\p_wT(w)\over(z-w)}+\ldots\cr
G^\a(z)\bar G^\b(w)=&{k\delta_{\a\b}\over(z-w)^3}+
	\half{ J^i(w)\sigma^i_{\a\b}\over(z-w)^2}+
		\half {T(w)\delta_{\a\b}\over(z-w)}+\ldots\cr
J^i(z)J^j(w)=&-\half{k\delta_{ij}\over (z-w)^2}+
	\e_{ijk}{J^k(w)\over(z-w)}+\ldots \cr
J^i(z)G^\a(w)=&-\half{\sigma^i_{\a\b}G^\b(w)\over(z-w})+\ldots ~.}}
The triplet of conserved charges is what is needed to construct the larger
spacetime supersymmetry algebra associated with a compactification down to
six, rather that four, dimensions. The $SU(2)$ Kac-Moody algebra is of
arbitrary level $k$, but we can see by comparison with \algone\ that the
central charge is constrained to be $6k$. Since
the level is constrained by unitarity to be integer, the only allowed values
of the central charge are $6,12,\ldots$~. Fortunately, $c=6$ is just what we
need, and this suggests that the N=4 algebra will be important role for us.

We will now show that a closer examination of the algebraic structure of the
wormhole conformal field theory reveals the existence of just the right
extended
supersymmetry. An important clue to understanding the structure of the
$(4,4)$ superconformal symmetry comes from the fact that there must be
{\it two} $SU(2)$ Kac-Moody symmetries:
The first is part of the standard N=4 superalgebra. This algebra contains the
energy-momentum tensor $T(z)$, four supercurrents $G^a(z)$ and three currents
$J^i(z)$ of conformal weight 1, which generate an $SU(2)$ Kac-Moody algebra
of a level tied to the conformal anomaly (in our case, level one).
The second is the $SU(2)$ Kac-Moody algebra of the Wess-Zumino-Witten
part of the wormhole conformal field theory. It has a general level $n$,
related to the area of the wormhole cross-section (or, equivalently, its
axion charge) and is clearly distinct from the N=4 $SU(2)$ Kac-Moody.
Since the superconformal algebra is quite tightly constrained, it is not
{\it a priori} obvious that such an $SU(2) \otimes SU(2)$ Kac-Moody is
compatible with N=4 supersymmetry and useful information, such as restrictions
on allowed values of the central charge, might be obtained by explicitly
constructing the algebra (assuming a consistent one to exist).
Quite remarkably, precisely the algebra we need has already been constructed by
Sevrin et. al. \sevrin\ , who discovered an alternate N=4 superalgebra,
containing an $SU(2)\otimes SU(2) \otimes U(1)$ Kac-Moody algebra, which had
been missed in previous attempts at a general classification of extended
superalgebras. In what follows\footnote*{This discussion was developed
in collaboration with E. Martinec}
we briefly summarize enough of their work to explain its significance for the
wormhole problem and, in particular, to verify the assertions made in section 4
about the r\^ole of (4,4) supersymmetry. In addition to establishing the
presence of a $(4,4)$ superconformal symmetry, this construction is a useful
starting point for studying the correspondence between the instanton moduli
space and perturbations of the superconformal field theory.

The construction discovered by Sevrin et.al. goes as follows: Start with the
bosonic WZW model for an $SU(2)\otimes U(1)$ group manifold (this is the
geometry of the wormhole if we let the radius of the $U(1)$ be infinite). The
conformal model contains four dimension-one Kac-Moody currents $J^a$
satisfying the usual KM algebra:
\eqn\kmcur{\eqalign{
J^0(z) J^0(w) =& -\half (z-w)^{-2} \cr
J^0(z) J^i(w) =& O( (z-w)^0 ) \cr
J^i(z) J^j(w) =& -\half n \delta^{ij}(z-w)^{-2}
	+\epsilon^{ijk}(z-w)^{-1} J^k(w) \cr}}
where $i=1,2,3$ indexes the currents of an $SU(2)$ algebra of level $n$ and
$J^0$ is the current of the $U(1)$ algebra. This is supersymmetrized by adding
a
set of four dimension-1/2 fields $\psi^a$ satisfying the {\it free} fermion
algebra
\eqn\fermalg{\psi^a(z)\psi^b(w)=-\half\delta^{ab}(z-w)^{-2}~~ }
(this is motivated by the arguments given earlier in this section that the
fermions in a supersymmetric wzw model are, modulo anomalies, free).

As usual, the Sugawara construction provides an energy-momentum tensor
\eqn\sugawar{T(z)=-J^0J^0-{1\over n+2}J^iJ^i-\partial\psi^a\psi^a}
with respect to which the fields $J^a$ ($\psi^a$) are primaries of weight
1 (1/2) and which has the expected $S_{wzw}$  conformal anomaly
\eqn\cexpct{c_{swzw}={3n\over n+2}+3=6(n+1)/(n+2)~.}
There is also a Sugawara-like construction of four real supersymmetry charges
$G^a$, with $a=0,..,3$ :
\eqn\suchrg{\eqalign{
G^0=&2[J^0\psi^0+(1/\sqrt{n+2})J^i\psi^i+(2/\sqrt{n+2})\psi^1\psi^2\psi^3] \cr
G^1=&2[J^0\psi^1+(1/\sqrt{n+2})(-J^1\psi^0+J^2\psi^3-J^3\psi^2)-
				(2/\sqrt{n+2})\psi^0\psi^2\psi^3]\cr}}
(plus cyclic expressions for $G^2$ and $G^3$). These supercharges could have
been packaged as a complex $I=1/2$ multiplet, as in \algfour\ . The operator
product expansion of these supercharges with themselves reads
\eqn\gope{\eqalign{
G^a(z) G^b(w)=&4{(n+1)\over(n+2)}\delta^{ab}(z-w)^{-3}+
            2\delta^{ab}T(w)(z-w)^{-1} \cr
-8[&{1\over n+2}\alpha^{+i}_{ab}A^+_i(w)+
           {n+1\over n+2}\alpha^{-i}_{ab}A^-_i(w)](z-w)^{-2}\cr
-4[&{1\over n+2}\alpha^{+i}_{ab}\partial A^+_i(w)+
           {n+1\over n+2}\alpha^{-i}_{ab}\partial A^-_i(w)](z-w)^{-1}\cr}}
where
\eqn\thooft{\alpha^{\pm i}_{ab}=\pm\delta^i_{[a}\delta^j_{b]}
           +\half\epsilon_{ijk} }
and
\eqn\sutoo{A^-_i=\psi^0\psi^i+\epsilon_{ijk}\psi^j\psi^k\qquad
A^+_i=-\psi^0\psi^i+\epsilon_{ijk}\psi^j\psi^k +J^i}
are commuting $SU(2)$ Kac-Moody algebras of levels 1 and n+1, respectively.
The c-number term (the central charge) and the term involving $T(z)$ are
obligatory in any higher-N superalgebra, while the terms involving dimension 1
operators are what differentiate the various possible extended superalgebras.
With further effort, one shows that the $G\cdot A^\pm$ OPE generates
combinations of $G^a$ and $\psi^a$ while the $G\cdot \psi$ OPE yields $A^\pm$
and $J^0$. No new operators appear in further iterations, so the complete
algebra generated by the supercharges contains just $T$ (dimension 2), $G^a$
(dimension 3/2), $A^i_\pm$ and $J^0$ (dimension 1) and $\psi^a$ (dimension
1/2).
The Kac-Moody algebra defined by the dimension 1 operators is evidently
$SU(2)\times SU(2)\times U(1)$ , which accords with our expectations derived
from the wormhole geometry.

The superalgebra whose construction we have outlined above is a particular
example of a one-parameter family of N=4 algebras dubbed the $A_\gamma$
algebras. The only problem with it is that the sigma model analysis of
extended supersymmetry (see for example \vnew ) makes quite clear that
the canonically defined supercharges and energy-momentum tensor must satisfy
the standard N=4 algebra, which closes on $T$, $G^a$ and a single level-one
$SU(2)$ Kac-Moody algebra $J^i$. The supercharges defined above obviously do
not have that property. However, if we `improve' them as follows
\eqn\improv{\tilde T=T-{1\over n+2}\partial J^0 \qquad
             \tilde G^a= G^a-{2\over n+2}\partial\psi^a ~~,}
we can show that $\tilde T$ , $\tilde G^a$ and $A^i_-$ (the level-one Kac-Moody
current) close on themselves and enjoy precisely the standard N=4
superalgebra. This says that the full algebra has the standard algebra as a
subalgebra, perhaps no great surprise.

This improvement has a simple physical interpretation: $J^0$ generates a $U(1)$
symmetry which can be regarded as a translation in a free coordinate $\rho$
(that is, we can write $J(z)\sim\p_z\rho(z)$ where $\rho$ is a free scalar
field). The original algebra \gope\ makes no reference to the dilaton and
corresponds physically to a constant dilaton field. It is well-known that,
if one turns on a dilaton which is {\it linear} in a free coordinate $\rho$,
this has the effect of adding a term proportional to $\p^2_z\phi\sim\p_zJ^0(z)$
to $T(z)$ and shifting the central charge of the superconformal algebra by
a constant. With a little care we can show that the linear dilaton implicit
in \improv\ is precisely what we obtained earlier in this section in our
discussion of the WZW-Feigin-Fuks conformal field theory of the wormhole.
This is a further piece of evidence that the improved energy-momentum tensor
$\tilde T$ is the physically relevant one. Now comes the miracle: $T$ is, in
any event, not physically acceptable because it has a central charge of
$6(n+1)/(n+2)$. The central charge of $\tilde T$ , however, can easily be
shown to be 6, precisely the required value!

This shows that there is an exact conformal field theory of just the right
central charge associated with the wormhole geometry and verifies the key
r\^ole of N=4 extended supersymmetry in establishing the physics of the model.
There are many fivebrane applications of this exact wormhole conformal field
theory which are just beginning to be worked out. Perhaps the most
interesting concern the vertex operators of excitations about the wormhole,
among which one must find the moduli of the exact solutions. In any event,
this line of argument shows that the dramatic consequences of (4,4)
superconformal symmetry, which we first extracted from perturbative
considerations, seem to have nonperturbative status.

\newsec{Conclusion}

In these lectures, we have constructed a special set of conformal field
theories which have the interpretation of soliton solutions of heterotic
string theory. We first constructed them as solutions of the leading
order in $\apm$ beta function conditions and then showed that, owing to
an extended worldsheet supersymmetry, the associated nonlinear sigma model
is an exact conformal field theory. It is the existence of an explicit and
exact conformal field theory associated with the soliton solution which
distinguishes the solution described here from previous attempts to
construct string theory solitons. There are several lines of inquiry which
can be pursued now that "exact" string solitons exist. One issue concerns
the mass of the soliton. In all previous discussions of string solitons,
the mass has been computed using the lowest-order spacetime effective action
\action\ and is therefore known only to lowest order in $\apm$.
It would obviously be desirable to know the mass exactly, but for that
one needs to develop a conformal field theory method of computing soliton
masses. Our exact soliton conformal field theory should provide a useful
laboratory for developing such methods. A second issue is
the question of stringy collective coordinates and their
semiclassical quantization. It should be an instructive challenge to
translate the well-known standard field theory physics of collective
coordinates into the string theory context. This is a nontrivial
matter because motion in collective coordinate space becomes motion in a
space of conformal field theories and it is a nontrivial matter to find
the action associated with such motions (and knowing the exact underlying
conformal field theories should help). Yet another question to pursue is
that of stringy black hole physics. We noted in Sect.~3 that our solitons
were similar to the extreme Reissner-Nordstrom black holes in the sense that,
while they have no singularity or event horizon, if one increases their mass
by any finite amount (while keeping the axion charge fixed), an event horizon
and a singularity (lying at a finite geodesic distance from any finite point)
will appear. Such black hole solitons can easily be created by scattering
some external particle on an extremal soliton and, by studying stringy
scattering theory about the extremal soliton, one should be able to explore,
in a controlled way, how a stringy black hole Hawking radiates and the nature
of the final state it approaches. These are quite difficult questions, but
having precise control of the underlying conformal field theory may allow
us to make progress on them. Perhaps it will be possible to report on
progress along these lines at the next Swieca Summer School.

\listrefs

\bye